# Analysis of Quantitative Angiography using Projection Foreshortening Correction and Injection Bias Removal

Parmita Mondal[1,2], Allison Shields[5], Mohammad Mahdi Shiraz Bhurwani[3], Kyle A Williams[1,2], Swetadri Vasan Setlur Nagesh[2,4], Adnan H Siddiqui[1,2,4], Ciprian N Ionita[1,2,3]

[1]Department of Biomedical Engineering, University at Buffalo, Buffalo, NY 14260
[2]Canon Stroke and Vascular Research Center, Buffalo, NY 14203
[3]Quantitative Angiographic Systems. Artificial Intelligence, Buffalo, NY 14203
[4]University at Buffalo Neurosurgery, Inc, Buffalo, NY 14203
[5]Yale School of Medicine, New Haven, CT 06510

## Abstract

**Background**: In neurovascular disease applications, 2D quantitative angiography (QA) based on digital subtraction angiography (DSA), is an intraoperative methodology used to assess disease severity and guide treatment. However, despite DSA's ability to produce detailed 2D projection images, the inherent dynamic 3D nature of blood flow and its temporal aspects can distort key hemodynamic parameters when reduced to 2D. This distortion is primarily due to biases such as projection-induced foreshortening and variability from manual contrast injection.

**Purpose:** This study aims to mitigate these biases and enhance QA analysis by applying a path-length correction (PLC) correction, followed by singular value decomposition (SVD)-based deconvolution, to angiograms obtained through both in-silico and in-vitro methods.

**Methods:** We utilized DSA data from in-silico and in-vitro patient-specific intracranial aneurysm models. To remove projection bias, PLC for various views were developed by co-registering the pre-existing 3D vascular geometry mask with the DSA projections, followed by ray tracing to determine paths across 3D vessel structures. These maps were used to normalize the logarithmic angiographic images, correcting for projection-induced foreshortening across different angles. Subsequently, we focused on eliminating injection bias by analyzing the corrected angiograms under varied projection views, injection rates, and flow conditions. Regions of interest at the aneurysm dome and inlet were placed to extract Time Density Curves for the lesion and the arterial input function, respectively. Using three standard SVD methodologies, we extracted the aneurysm Impulse Response function (IRF) and its associated parameters Peak Height (PH$_{IRF}$), Area Under the Curve (AUC$_{IRF}$), and Mean Transit Time (MTT).

**Results:** Our findings revealed that projection and injection parameters significantly affect key quantitative angiographic parameters such as PH$_{IRF}$, AUC$_{IRF}$, and MTT. Despite these variations, our approach utilizing PLC allowed by SVD-based deconvolution consistently nearly eliminated these effects across both in-silico and in-vitro settings, yielding stable and reliable measurements which were correlated only with the hemodynamic conditions.

**Conclusion:** Our methodology employing PLC and SVD-based deconvolution ensures reliable quantitative angiographic measurements across varying conditions, supporting consistent assessments of disease severity and treatment efficacy. This approach significantly enhances intrapatient and intraprocedural reliability in neurovascular diagnostics.

# 1  Introduction

Intracranial aneurysms (IAs) are balloon-like malformations that, if left untreated, could lead to subarachnoid hemorrhage (SAH), a type of stroke with high morbidity and mortality.[1] Hemodynamics plays a critical role in the IA's growth and rupture, as well as, the post-treatment healing progression. Treatments predominantly aim to either eliminate the aneurysm through surgical clipping or to reduce rupture risk and promote healing via endovascular techniques that alter hemodynamics, such as coils and flow diverters. Given the aneurysm's intracranial location and the vasculature's intricate, tortuous nature, direct intraoperative hemodynamic assessments using current technologies like ultrasound or through endovascular devices are impractical. Consequently, interventionalists rely on a qualitative evaluation based on the intra-arterial flow of the contrast agent via digital subtraction angiography (DSA).[2] For nearly three decades, digital subtraction angiography (DSA) has remained the primary imaging modality during these procedures, providing essential visualization of blood vessels and aneurysms.[3]

More recently, two-dimensional quantitative angiography (2D-QA) has been proposed as an improvement over traditional qualitative angiography, i.e., visualization of DSA loops, demonstrating significant potential in predicting treatment outcomes. The 2D-QA method involves analyzing contrast behavior by synthesizing time-density curves (TDCs) from specified regions of interest (ROIs), as illustrated in Figure 1, and extracting key parameters from these curves, which are then used in prognosis algorithms. [4] Despite its promise, 2D-QA still faces significant limitations, primarily due to hand injection variability and foreshortening bias.

Hand injection variability introduces significant bias due to inconsistencies in manual contrast injections. Many prior investigations have explored normalization techniques and deconvolutions to reduce discrepancies arising from variations in injection force and duration. To mitigate hand injection variability, various techniques, including normalization to arterial input function (AIF) as well as deconvolution have been proposed. [5] These approaches have shown to improve prognosis algorithms and nearly remove the injection variability. [6]

Projection foreshortening bias arises from DSA's intrinsic limitation in depth integration, where overlapping vessel structures and different X-ray path lengths contribute to potentially misguiding TDC readings. [7] This depth integration leads to a blending of flow information, resulting in distorted hemodynamic parameters. [8] Imaging biomarkers, including peak height (PH), area under the curve (AUC), and mean transit time (MTT), are often misrepresented. As shown in Figure 1, where a simulated biplane is presented, foreshortening affects the TDCs extracted from both the aneurysm dome and the AIF. Assuming the same angiographic run from two projections, the amount of contrast in the aneurysm and vessel is identical; however, the recorded gray values at the detector are directly correlated with the X-ray path length through the aneurysm. Previous work proposed a baseline correction method, normalizing each view to pretreatment angiograms to reduce pathlength bias However, this approach poses challenges, especially when C-arm positioning changes between pre and post-treatment DSA runs.

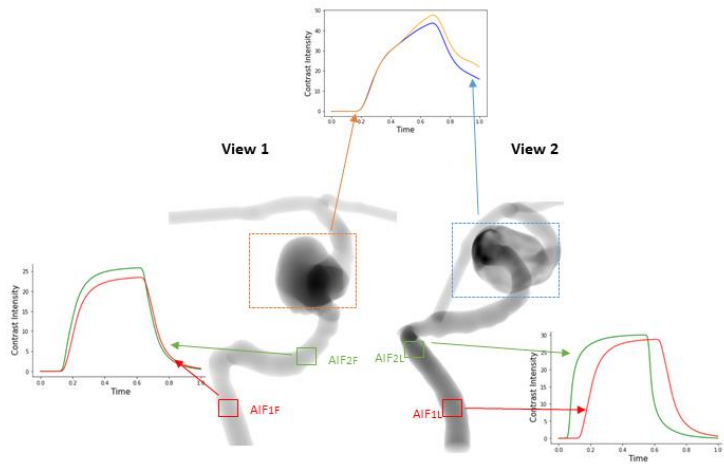

**Figure 1.** Different orientation views of the same virtual angiogram are generated using computational fluid dynamics (CFD). We drew ROI of aneurysm dome for two different views, to obtain two different TDCs. We draw ROI at the two different arterial input functions (AIFs), to demonstrate the TDCs generated from them.

To address these challenges, we propose investigating a method for foreshortening correction that assumes the availability of a 3D volume. This is clinically feasible, as modern angiographic studies typically include a C-arm spin and cone beam reconstruction of the lesion of interest. We will conduct in-silico and in-vitro studies to demonstrate the feasibility and assess the accuracy of this correction approach.

## 2 Materials and Methods

### 2.1 In-silico experiment

Data analysis was approved by our institutional review board. All data for this study were acquired at Gates Vascular Institute in Buffalo, NY. The segmentation of the models began with the Vitrea 3D station (Vital Images, Inc., Minnetonka, MN), where the focus was on the aneurysm dome and the inlet ROI. Subsequent refinements were applied to the exported stereolithographic (STL) files using Autodesk Meshmixer (Autodesk Inc., San Francisco, CA) to enhance the smoothness of the arterial wall. [9] To create high-resolution meshes that accurately represent the arterial geometry, the refined STL files underwent mesh generation using ICEM (Ansys Inc., Canonsburg, PA). These meshes were imported into Fluent (Ansys Inc., Canonsburg, PA), where steady-state laminar flow was simulated by solving the Navier-Stokes equations while enforcing inlet velocity functions and a zero-pressure condition at the vessel outlet. [8,10] A parabolic velocity function was defined at each inlet cross-section, with mean velocities of 0.25 m/s, 0.35 m/s, and 0.45 m/s. The simulations met the convergence criteria of 1e−6 at each timestep, and the SIMPLE scheme was implemented with a second-order formulation.

Virtual angiograms were generated by labeling all particles entering the inlet during specified time intervals and tracing their trajectories through the vascular network. To simulate different durations of iodinated injections, the labeling was activated for durations of 0.5s, 1.0s, 1.5s, and 2.0s, respectively. For each simulation, the concentration was calculated at every volume element in the meshed geometry. This approach provided a time-resolved, 3D angiogram with a temporal resolution of 1 millisecond. The distribution of contrast was then exported as a point cloud, where each value represents the concentration of labeled particles in each volumetric mesh. For further processing and to generate 2D angiographic sequences, the concentration point clouds were imported into MATLAB (R2021b, Natick, MA). Using this environment, rotations were applied to the 3D simulated angiograms to achieve optimal 2D projection views of the aneurysm dome. For the synthesis of DSA runs, temporal integration over the 3D virtual angiogram was used to simulate a 10-millisecond frame, followed by a cone beam forward projection using the ASTRA toolbox, implemented in Python (Python 3.10, Wilmington, DE).

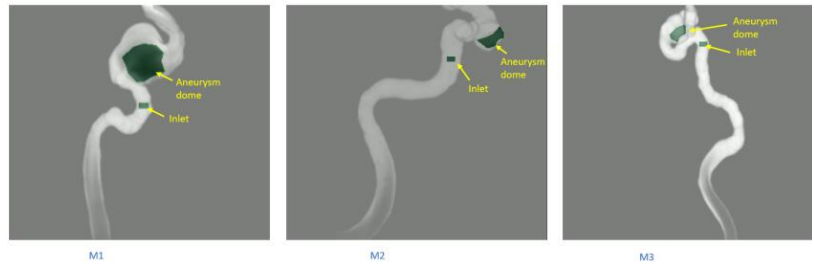

**Figure 2. Snapshot of PLC for one frame in the angiographic sequence for three different aneurysms.** Yellow arrows show the aneurysm dome and inlet ROI that will be used to extract the TDCs for data analysis.

In-silico path length correction (PLC) is straightforward, particularly because the vascular geometry used to generate the virtual angiograms is already aligned with the 2D angiogram. In the process of creating a path-length-corrected projection image, all rays contributing to the image formation are summed and weighted based on each ray's intersection length with the vessel (i.e., rays passing through a longer portion of the vessel have a greater impact on the final projection image compared to those passing through shorter segments). [11] Using this approach, we generated a weight matrix, which provides the intersection of rays with the vessel in centimeters. [8].The contrast data from the virtual angiogram were then divided by this weight matrix to obtain the path-length-corrected virtual angiogram. These steps were repeated for all aneurysm cases, flow conditions, and injection simulations.

## 2.2 Experimental setup for in-vitro analysis

Our experimental setup was modeled after the methodology outlined by Ionita et al., utilizing a patient-specific internal common carotid artery phantom to simulate realistic physiological conditions[3]. A cardiovascular waveform pump (Model 1407, Harvard Apparatus, MA) operated at 70 pulses per minute, with a systole/diastole ratio of 30/70 and a stroke volume of 20 cc. A damper was included to reduce pulse wave effects from the pump. A contrast injector catheter and pressure transducer probe were placed between the damper and phantom for precise contrast injection and monitoring. Iodine-based contrast agent was injected via a 7Fr catheter connected to an automatic injector.

Fluid velocity measurements taken at the main branch using an ultrasound probe and flowmeter showed an average flow of 1.91 l/min and carotid velocity of 63.3 cm/s. Used contrast was collected in a separate container to maintain system integrity. Bolus injection volumes and durations were carefully controlled and varied to assess their effect on imaging. DSA images were captured at 15 frames per second using a Canon Infinix bi-plane system, Figure 3. A 5 ml bolus was injected at four rates: 5 ml/s, 10 ml/s, 15 ml/s, and 20 ml/s, corresponding to injection durations of 1.0 s, 0.5 s, 0.33 s, and 0.25 s, respectively.

For the in-vitro PLC, we acquired rotational DSAs and performed 3D reconstructions using the Feldkamp-Davis-Kress (FDK) algorithm with Parker weighting, Figure 3. After reconstruction, DSA DICOM metadata was used to roughly align the segmented 3D reconstruction with the DSA projections. This initial alignment was followed by a cone-beam forward projection in ASTRA, utilizing the DICOM metadata for table position and source-to-imager distance. For co-registration of the projected mask with an averaged DSA frame, we applied an initial affine transformation to align the forward-projected 3D reconstruction with the arteries. This coarse alignment was achieved using the SimpleITK library's Elastix, with Advanced Mattes Mutual Information as the metric and Quasi Newton LBFGS as the optimizer. The affine transformation was configured with five resolutions, and automatic scale estimation was enabled to dynamically adjust the transformation scales. Following the affine transformation, the alignment was refined using B-spline deformable registration to account for anatomical differences. The B-spline registration, also implemented using Elastix Image Filter, was configured with four resolutions and a grid spacing schedule that progressively refined from 128 to 0.01 for precise alignment, Figure 3. Elastic regularization ensured a balance between smoothness and flexibility during transformation. Once co-registration was achieved, the transformation was applied to the PLC 2D matrix generated as described earlier in the in-silico process. Each angiogram was then divided by the path-length matrices for both frontal and lateral views, Figure 3.

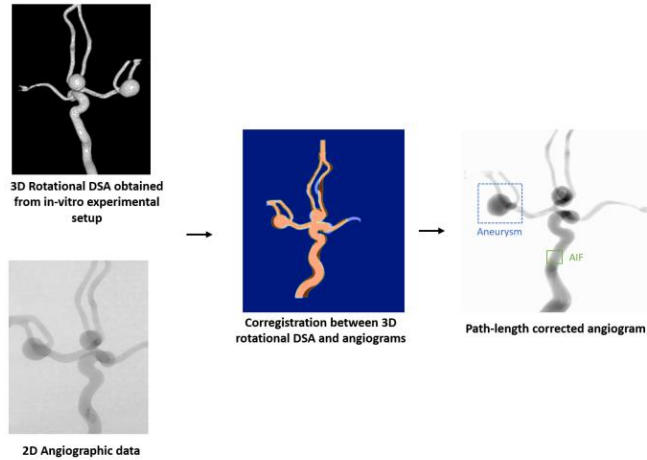

**Figure 3.** Demonstrates the corregistration between 3D rotational DSA and 2D angiograms, followed by path-length correction, on in-vitro acquired angiograms.

## 2.3 Quantitative angiographic parameters analysis using deconvolution methods

To mitigate the impact of injection variability on QA parameters, we applied singular value decomposition (SVD)-based deconvolution techniques to the path-length corrected in-silico and in-vitro angiograms. Specifically, we adapted three SVD variants commonly used in perfusion analysis: standard SVD with Tikhonov regularization (sSVD), block-circulant SVD (bSVD), and oscillation index SVD (oSVD). [6]

In these methods, the arterial input function (AIF), represented by the inlet TDC, is used to deconvolve the aneurysm response. For sSVD, the AIF is arranged into a Toeplitz matrix, while for bSVD and oSVD, it is transformed into a block-circulant matrix. [12,13] The deconvolution yields the impulse response function (IRF) of the aneurysm dome, from which we derive key QA parameters. To reduce the effects of noise, we applied truncation to the singular values in the SVD process. The optimal truncation threshold ($SVD_{trunc}$) was determined empirically, retaining singular values above a certain percentage of the maximum value.

From the obtained IRF, we evaluated the peak height ($PH_{IRF}$), area under the curve ($AUC_{IRF}$), and mean transit time (MTT). $PH_{IRF}$ is the maximum value of the IRF, $AUC_{IRF}$ is the integral of the IRF, and MTT was calculated as $AUC_{IRF}$ divided by $PH_{IRF}$.[6] This process was repeated for each deconvolution method, allowing us to assess the effectiveness of each in stabilizing QA parameters against injection variability.

## 2.4 Data Analysis

Using the in-silico and in-vitro experiments described above, we aim to validate two key hypotheses using PLC and SVD methods. First, we hypothesize that PLC eliminates view bias in arterial input TDCs. To test this, we selected two regions of interest (AIFs) at the arterial inlet for the frontal proximal and distal ($AIF_{1F}$, $AIF_{2F}$) and lateral ($AIF_{1L}$, $AIF_{2L}$) views, as shown in Figure 4. These ROIs correspond to the same anatomical locations in different views, as well as nearby ROIs within the same view but slightly shifted to sample different regions of the artery. If the PLC method is correct, the arterial input function (AIF) should be invariant across views and should not change when shifting the inlet ROI by a small distance. We generated TDCs for each ROI and computed the root mean square error (RMSE) between them. A significant reduction in RMSE after PLC, would confirm that PLC effectively removes view bias and corrects for foreshortening errors.

Second, we hypothesize that PLC combined with SVD eliminates injection bias. Specifically, after applying PLC and SVD, the mean transit time (MTT), which measures the average time a particle spends in a given ROI, should be independent of the injection duration and reflect flow dynamics within the aneurysm. We analyzed the MTT across varying injection durations and flow rates, expecting the MTT to remain consistent if the injection bias was successfully removed. By examining the MTT slope under these conditions, we could confirm that PLC and SVD minimize injection-related variability. In addition, we investigated how different SVD algorithms might affect the results.

## 3   Results

We systematically evaluated the efficacy of PLC across three distinct in-silico aneurysm models, each subjected to three different inlet velocities, 0.25m/s, 0.35m/s and 0.45m/s, and four injection durations, 0.5s, 1s, 1.5s and 2s, analyzed from two angiographic views using three singular SVD methodologies. Complementarily, our in-vitro analysis was conducted on a patient-specific model, assessing two views and four varying injection durations.

In Figure 4, we illustrate the impact of PLC on the selection of AIFs and the subsequent derivation of aneurysm IRFs across both in-silico and in-vitro studies. Initially, AIFs derived from different ROIs without PLC exhibited substantial variability, as demonstrated by distinct TDC profiles leading to divergent IRF shapes and parameters. Upon implementing PLC, the variability in AIF selection was reduced, resulting in location-invariant AIFs and uniformly derived IRFs from the aneurysm dome, regardless of the ROI location.

These results were quantitatively supported by a significant reduction in RMSE between TDCs from different ROIs post-PLC application. Table 1 displays RMSE values both before and after the implementation of PLC. Prior to PLC, significant RMSE values are evident across different AIFs, as shown in the 'Before PLC' tables for both in-silico and in-vitro models. These high RMSE values highlight the significant inconsistencies and variability in the AIFs due to projection-induced biases. Post-PLC implementation, there is a clear reduction in the RMSEs for all AIFs across all tested scenarios, as illustrated in the 'After PLC' tables.

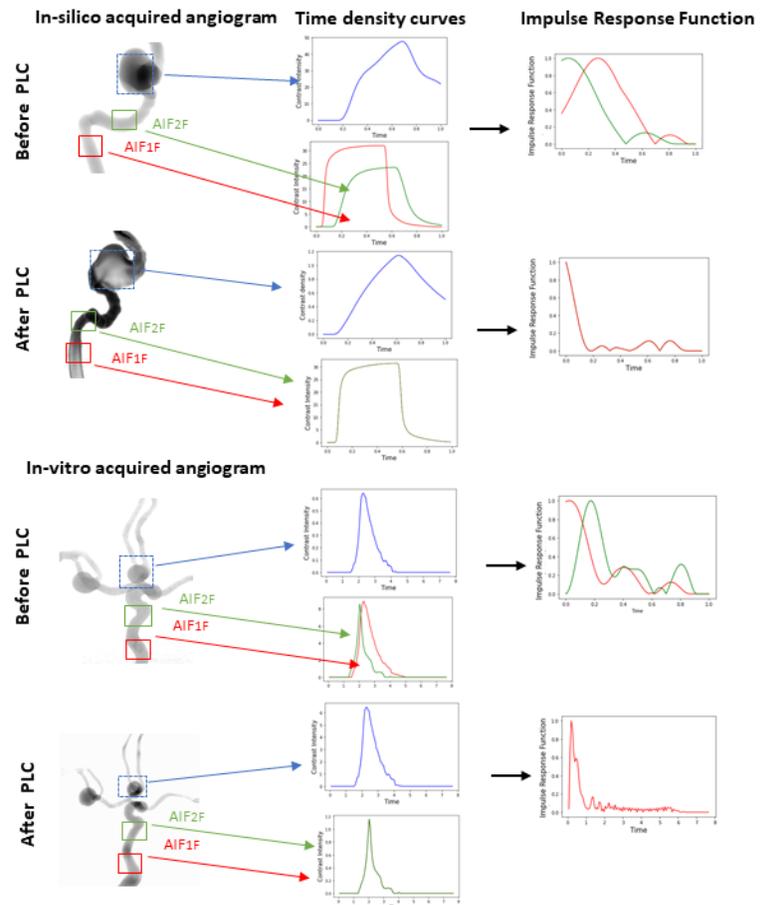

**Figure 4.** Impact of path-length correction (PLC) on in-silico and in-vitro angiograms, demonstrated through alterations in arterial input functions (AIFs) and impulse response functions (IRFs). The top two rows display in-silico angiograms pre- and post-PLC, while the bottom rows show in-vitro results under the same conditions. Corresponding time density curves (TDCs) and IRFs highlight variability with different AIFs before PLC and uniformity after applying PLC, confirming its effectiveness in stabilizing angiographic analysis.

MTT analysis for one in silico model and the in-vitro models are shown in Figures 5 and 6 respectively. Figure 5, focusing on in-silico models, plots MTT against injection durations for different SVD methods at three inlet velocities (0.25 m/sec, 0.35 m/sec, 0.45 m/sec), and Figure 6 for in-vitro models compares MTT changes from frontal and lateral views under the same flow conditions. In both figures, the trend lines

**Table 1**: RMSE values between TDCs from different AIFs in both in-silico and in-vitro models before and after PLC. For in-silico angiograms, RMSEs are averaged across three aneurysm models, comparing $AIF_F$ for view 1 and $AIF_L$ for view 2. In-vitro data, derived from a single model, also compares RMSEs for similar locations and views pre- and post-PLC.

**Before PLC**

**In-silico**

|  | $AIF_{1F}$ | $AIF_{2F}$ | $AIF_{1L}$ | $AIF_{2L}$ |
|---|---|---|---|---|
| $AIF_F$ | 0 | 2.27 ± 0.89 | 8.41 ± 4.77 | 10.70 ± 2.17 |
| $AIF_{2F}$ | 2.27 ± 0.89 | 0 | 7.98 ± 4.53 | 10.02 ± 2.07 |
| $AIF_{1L}$ | 8.41 ± 4.77 | 7.98 ± 4.53 | 0 | 6.53 ± 1.74 |
| $AIF_{2L}$ | 10.70 ± 2.17 | 10.02 ± 2.07 | 6.53 ± 1.74 | 0 |

**After PLC**

|  | $AIF_{1F}$ | $AIF_{2F}$ | $AIF_{1L}$ | $AIF_{2L}$ |
|---|---|---|---|---|
| $AIF_F$ | 0 | 0.07 ± 0.01 | 0.08 ± 0.01 | 0.08 ± 0.01 |
| $AIF_{2F}$ | 0.07 ± 0.01 | 0 | 0.10 ± 0.02 | 0.09 ± 0.03 |
| $AIF_{1L}$ | 0.08 ± 0.01 | 0.10 ± 0.02 | 0 | 0.03 ± 0.01 |
| $AIF_{2L}$ | 0.08 ± 0.01 | 0.09 ± 0.03 | 0.03 ± 0.01 | 0 |

**In-vitro**

|  | $AIF_{1F}$ | $AIF_{2F}$ | $AIF_{1L}$ | $AIF_{2L}$ |
|---|---|---|---|---|
| $AIF_F$ | 0 | 0.12 | 0.46 | 0.58 |
| $AIF_{2F}$ | 0.12 | 0 | 0.51 | 0.65 |
| $AIF_{1L}$ | 0.462 | 0.51 | 0 | 0.31 |
| $AIF_{2L}$ | 0.589 | 0.65 | 0.31 | 0 |

|  | $AIF_{1F}$ | $AIF_{2F}$ | $AIF_{1L}$ | $AIF_{2L}$ |
|---|---|---|---|---|
| $AIF_F$ | 0 | 0.02 | 0.06 | 0.08 |
| $AIF_{2F}$ | 0.02 | 0 | 0.07 | 0.08 |
| $AIF_{1L}$ | 0.06 | 0.07 | 0 | 0.03 |
| $AIF_{2L}$ | 0.08 | 0.08 | 0.03 | 0 |

colored in orange represent scenarios without PLC and deconvolution, showing a significant MTT dependency on injection duration. The gray lines show the effects when employing SVD methodologies alone, which show already significant reduction in variability. Finally, the blue lines, denote the implementation of both PLC and SVD, exhibit the flattest slopes, highlighting a minimized dependency on injection parameters.

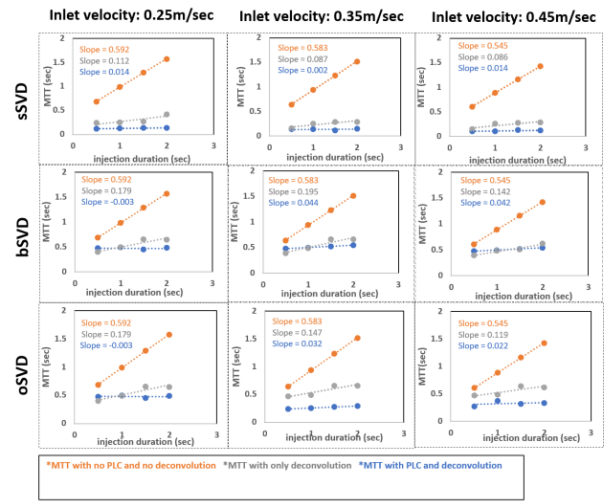

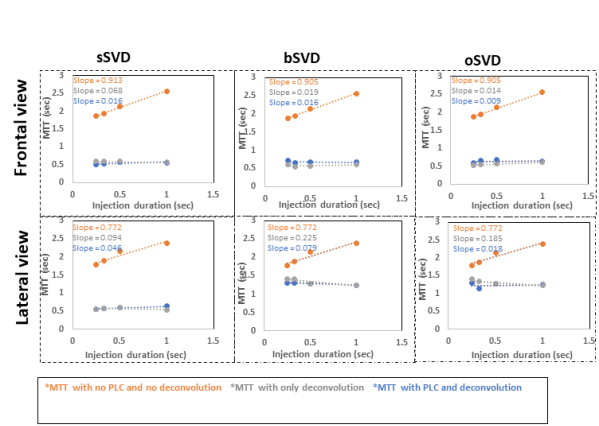

**Figure 5.** Plot of MTT after the implementation of sSVD (A), bSVD (B) and oSVD (C) is shown. The first column is at inlet velocity 0.25m/sec, the second column is at inlet velocity 0.35m/sec and the last column is at inlet velocity 0.45m/sec, respectively.

**Figure 6.** Plot of MTT after the implementation of path-length correction and sSVD, bSVD and oSVD is shown, for in-vitro acquired angiograms, for frontal (A) and lateral (B) view.

Analysis of MTT slope results for the three in-silico models and three velocities, analyzed using different SVD variants with PLC were: 0.015 ± 0.017, 0.014 ± 0.032 and 0.013 ± 0.025, for sSVD, bSVD and oSVD respectively. Similarly, for the in-vitro studies using two views the MTT slopes were: 0.031 ± 0.015, 0.047 ± 0.031and 0.013 ± 0.004, for the same SVD techniques.

## 4   Discussion

In this study, we aimed to address two major limitations in quantitative angiography (QA) for intracranial aneurysm assessment: projection-induced foreshortening bias and variability from manual contrast injection. By implementing a path-length correction (PLC) method followed by singular value decomposition (SVD)-based deconvolution on both in-silico and in-vitro angiograms, we demonstrated significant improvements in the accuracy and reliability of QA parameters.

Our findings indicate that PLC effectively reduces the foreshortening bias inherent in 2D DSA images. Before PLC, time-density curves (TDCs) extracted from different regions of interest (ROIs) exhibited substantial variability due to differences in projection angles and tortuous vessel structures. This variability was quantitatively evidenced by high root mean square error (RMSE) values between TDCs from closely placed ROIs over the main feeding artery. After applying PLC, the RMSE values significantly decreased, confirming that PLC minimizes view-dependent discrepancies and standardizes the arterial input function (AIF) across different projections. This supports our first hypothesis that PLC eliminates view bias in arterial input TDCs.

Furthermore, the application of SVD-based deconvolution methods to the PLC-corrected angiograms effectively mitigated the impact of injection variability on QA parameters[14]. The mean transit time (MTT), which is sensitive to variations in injection rates and durations, showed a strong dependency on these factors before correction. However, after applying PLC and SVD, the MTT values became more consistent across varying injection conditions, as evidenced by the flattened slopes in Figures 5 and 6. This validates our second hypothesis that PLC combined with SVD eliminates injection bias.

Among the SVD variants tested, all methods (sSVD[15,16], bSVD[13], oSVD[17]) demonstrated improvements in stabilizing QA parameters. However, subtle differences were observed in their performance. The sSVD method with Tikhonov regularization provided a good balance between noise suppression and preservation of hemodynamic information. The bSVD and oSVD methods, which use block-circulant matrices and incorporate oscillation indices, respectively, also showed efficacy in reducing injection-related variability. The choice of SVD variant may depend on specific clinical requirements and computational considerations. Our results have significant implications for neurovascular diagnostics and treatment planning[18-20]. By correcting for projection and injection biases, clinicians can obtain more reliable and accurate measurements of hemodynamic parameters such as $PH_{IRF}$ $AUC_{IRF}$, and the MTT derived by dividing the two term. This can enhance the assessment of disease severity, improve the prediction of treatment outcomes, and potentially lead to more personalized therapeutic strategies.

Despite these promising results, our study has limitations. The in-silico models, while providing controlled conditions, may not capture all the complexities of in vivo hemodynamics, such as x-ray parameters effect, blood flow and patient-specific anatomical variations. Similarly, the in-vitro experiments, although utilizing patient-specific phantoms, do not fully replicate the biological environment, including factors like vessel elasticity and blood viscosity variations. Additionally, the application of PLC and SVD methods requires the availability of high-quality 3D imaging data for accurate co-registration and path-length mapping, which is available in most clinical setting but may not always be available.

Future research should focus on validating these methods in clinical studies involving patient data to assess their practical utility and robustness in real-world scenarios. Moreover, exploring the integration of PLC and SVD techniques with advanced reconstructions techniques such as epipolar reconstruction, could further enhance the understanding of intracranial aneurysm hemodynamics.

# 5 Conclusion

In conclusion, our study demonstrates that the combination of path-length correction and SVD-based deconvolution significantly enhances the reliability of quantitative angiographic measurements by mitigating projection and injection biases. This methodological advancement holds potential for improving intraprocedural assessments and optimizing treatment strategies for patients with intracranial aneurysms.

# 6 Acknowledgements

This work was supported by QAS.AI and NSF STTR Award # 2111865

# 7 Conflict of interest Statement

The authors have no relevant conflicts of interest to disclose.